\documentclass[review]{elsarticle}
\graphicspath{ {./figures/} }
\usepackage{hyperref}
\usepackage{float}
\usepackage{verbatim} 
\usepackage{apalike}
\usepackage{algpseudocode}
\usepackage{amsmath}
\usepackage{amsfonts}
\usepackage{mathtools}
\usepackage{amssymb}
\restylefloat{figure}
\restylefloat{table}
\usepackage{xcolor,color}
\usepackage{multirow}
\usepackage{colortbl}
\usepackage{booktabs}

\journal{Expert Systems with Applications}

\biboptions{authoryear}

\begin{document}
\begin{frontmatter}

\begin{titlepage}
\begin{center}
\vspace*{1cm}

\textbf{ \large {Modeling IoT Traffic Patterns: Insights from a Statistical Analysis of an MTC Dataset}}

\vspace{1.5cm}

David E. Ruiz-Guirola$^{a}$ (david.ruizguirola@oulu.fi), Onel L. A. L\'{o}pez$^a$ (onel.alcarazlopez@oulu.fi), Samuel Montejo-S\'{a}nchez$^b$ (smontejo@utem.cl) \\

\hspace{10pt}

\begin{flushleft}
\small  
$^a$ P.O.Box 8000, FI-90014 University of Oulu, Oulu, Finland \\
$^b$ Programa Institucional de Fomento a la Investigaci\'{o}n, Desarrollo e Innovaci\'{o}n, Universidad Tecnol\'{o}gica Metropolitana, 8940577 Santiago, Chile \\

\vspace{1cm}
\textbf{Corresponding Author:} \\
David E. Ruiz-Guirola \\
P.O.Box 8000, FI-90014 University of Oulu, Oulu, Finland\\
Tel: (+358) 294 48 0000 \\
Email: David.RuizGuirola@oulu.fi

\end{flushleft}        
\end{center}
\end{titlepage}

\title{Modeling IoT Traffic Patterns: Insights from a Statistical Analysis of an MTC Dataset}

\author[label1]{David E. Ruiz-Guirola \corref{cor1}}
\ead{david.ruizguirola@oulu.fi}

\author[label1]{Onel L. A. L\'{o}pez}
\ead{onel.alcarazlopez@oulu.fi}

\author[label2]{Samuel Montejo-S\'{a}nchez }
\ead{smontejo@utem.cl}

\cortext[cor1]{Corresponding author.}
\address[label1]{P.O.Box 8000, FI-90014 University of Oulu, Oulu, Finland}
\address[label2]{Programa Institucional de Fomento a la Investigaci\'{o}n, Desarrollo e Innovaci\'{o}n, Universidad Tecnol\'{o}gica Metropolitana, 8940577 Santiago, Chile}

\begin{abstract}
The Internet-of-Things (IoT) 
is rapidly expanding, connecting numerous devices and becoming integral to our daily lives. 
As this occurs, 
ensuring efficient traffic management 
becomes crucial. 
Effective IoT traffic management 
requires 
modeling and predicting intrincate machine-type communication (MTC) dynamics, for which machine-learning (ML) techniques are certainly appealing. 
However, obtaining comprehensive and high-quality datasets, along with accessible platforms for reproducing ML-based predictions, continues to impede the research progress. 
In this paper, we aim to fill this gap by characterizing the Smart Campus MTC dataset provided by the University of Oulu. 
Specifically, we 
perform a comprehensive statistical analysis of the MTC traffic utilizing goodness-of-fit tests, including well-established tests such as Kolmogorov-Smirnov, Anderson-Darling, chi-squared and root mean square error. 
The analysis centers on examining and evaluating three models that accurately represent the two most significant MTC traffic types: periodic updating and event-driven, which are also identified from the dataset. 
The results demonstrate that the models accurately characterize the traffic patterns. The Poisson point process model exhibits the best fit 
for event-driven patterns with errors below 11\%, while the quasi-periodic model fits accurately the periodic updating traffic with errors below 7\%.  

\end{abstract}

\begin{keyword}
Goodness-of-fit, MTC traffic, ML-based predictor, Smart Campus.
\end{keyword}

\end{frontmatter}

\section{Introduction}
\label{introduction}

Machine-type communication (MTC), a crucial component of the Internet of Things (IoT), is a specialized form of communication that focuses on machine-to-machine communication without any human intervention \citep{aldahiri2021trends}. 
As MTC networks continue to grow rapidly, managing and optimizing resources has become a crucial challenge for ensuring scalability. Additionally, the low-power/complexity of the MTC devices (MTD) present another challenge in terms of data management and network availability. These factors depend on the limited battery lifetime of the devices and their ability to implement algorithms, making energy efficiency and optimization critical enablers for future MTC networks \citep{shafiq2020corrauc}. 


Characterizing and modeling MTC traffic is crucial for optimizing wireless IoT networks by tailoring management strategies to specific application needs~\citep{sharma2019toward,sharma2018distributed}. With this, significant energy savings may be achieved, which is crucial due to the limited battery lifespan inherent in IoT networks, thereby improving network efficiency and scalability~\cite{shafiq2020corrauc}. 
This may be enabled by the exploitation of accurate machine learning (ML)-based traffic predictors~\citep{kato2016deep}. For example, idle channel monitoring is responsible for wasting over half of the energy consumed in these networks~\citep{mughees2020towards} while the results in~\citep{IoT_FWuS} indicated that up to 38\% of the consumed energy could be saved by exploiting a prediction method when using sleep modes like wake-up radio or discontinuous reception. 
Unfortunately, the use of ML requires numerous labeled data obtained  from extensive, large-scale dataset~\citep{aldahiri2021trends}. 

A significant stage of the ML process is the data analysis, which can be a difficult task. A thorough understanding of a dataset is crucial to ML since features that are misunderstood or incorrectly generated can lead to inaccurate models~\citep{peterson2021review,aldahiri2021trends}. 
One of the most significant drawbacks of developing and validating detection schemes is the lack of data sets~\citep{himeur2021artificial,sousa2020wildfire}.  Many ML models are evaluated on a small number of examples rather than comprehensive data sets and are not accessible to the research community.
Meanwhile, developing accurate ML models usually require exploiting extensive and high-quality data sets~\citep{fourati2021survey}.  

Additionally, the lack of platforms to test the results of existing ML-based predictions 
limits the performance comparison of existing algorithms and makes it difficult to understand the current state of the art~\citep{himeur2021artificial}. Moreover, the management of large amounts of data usually requires intense operations that are real-time oriented and have a high demand for computational, memory, and energy resources~\citep{fourati2021survey}.
For this reason, academia and industry are creating and developing infrastructures to generate their own datasets for research, as in~\citep{koumaras20185genesis}. For example, ~\citep{le2018applying} applies big data, ML, and network function virtualization to create a practical framework for traffic management across different cell types (GSM, 3G, and 4G). Authors in~\citep{de2018importance} present a dataset of 250,000 over-the-air transmissions with heterogeneous transceiver hardware and co-channel interference, while \citep{kotz2005crawdad} has an open archive with wireless trace data from numerous contributors. Such datasets constitute an opportunity for the development and testing of ML algorithms. 
Despite recent advances in current literature that utilize deep learning architectures, 
there is still a lack of understanding of the limitations of these models in real scenarios~\citep{sousa2020wildfire}. 
Synthetic data generation techniques may be used to address this issue. This involves creating artificial data sets to simulate real-world scenarios, and can be used to test and validate ML models~\citep{IoT_FWuS}. Specifically, spatio-temporal models are developed in \citep{emara2020spatiotemporal} to study the spatial and temporal dynamics of MTC networks. In the field of mathematical modeling, the authors in~\citep{smiesko2023markov} use two-state Markov-modulated On-Off processes to represent different types of IP flows and enable the construction of relatively simple Markov queuing systems for network nodes. Meanwhile, in \citep{laner2013traffic}, the authors propose a coupled Markov modulated Poisson process for the source traffic model. 
From the ML perspective, the authors in~\citep{ruiz2023performance} assess the performance of different ML methods for predicting Poisson and quasi-periodic MTC traffic in terms of accuracy and computational cost. The paper offers a unified framework for comparing ML predictors under different traffic models, aiding the selection of appropriate models for energy-efficient MTC applications. 


Nonetheless, an ongoing disparity exists among the traffic models utilized in various studies when using ML.  
Generally, there are two main types of traffic models: source traffic models and aggregated traffic models. Source traffic models focus on understanding and replicating the behavior of individual data sources, such as video streams or data transfers where each source is considered independently. On the other hand, aggregated traffic models deal with the combined traffic behavior of the entire network or a large segment of it, considering the combined behaviors of various sources instead of focusing on individual sources.  
Achieving higher prediction accuracy often demands source traffic models, while opting for aggregated traffic models can lead to reduced complexity~\citep{navarro2020survey}. 
Reliable and comprehensive traffic models are crucial in reducing the cost of training ML predictors using real data. 
However, demonstrating the validity of these models for such scenarios by studying 
MTC IoT databases with real data is still an aspect that requires extensive research for different MTC scenarios. 

In this paper, we aim to fill the existing research gap by conducting extensive studies using real data from MTC IoT database, ultimately contributing to the improved understanding and application of these models in practical scenarios when using ML algorithms. 
Specifically, this paper focuses on characterizing the Smart Campus dataset from~\citep{realdata} and conducting a statistical analysis using goodness-of-fit tests like Kolmogorov-Smirnov (K-S), Anderson-Darling (A-D), and chi-squared (C-S). 
The Smart Campus dataset is a collection of data obtained from low-power sensors that mirrors the communication patterns inherent in MTC environments, providing insights into real-world MTC traffic behaviors. 
Our analysis here helps identify and study the three models that best represent the two most relevant MTC traffic types: (i) event-driven traffic patterns and (ii) periodic update traffic patterns, including those recommended by \citep{3gpp2011study}. Furthermore, we compare and analyze the most suitable traffic models for simulating real-world scenarios corresponding to each traffic type by assessing the goodness-of-fit of our proposal and other well-known models. The models' accuracy is compared, and a detailed analysis of their goodness-of-fit is presented to demonstrate the validity of these models for various MTC scenarios, using well-known tests such as K-S and root mean square error (RMSE). The results show that the Poisson point process (PPP) model provides the best fit for event-driven patterns, while the quasi-periodic model exhibits the best fit for periodic updating.



The rest of the paper is organized as follows. Section II presents and characterizes the dataset. Section III describes the statistical analysis of the dataset and presents the  MTC traffic models proposed: PPP and quasi-periodic. Meanwhile, Section IV introduces the performance metrics and Section V presents the comparison results between the dataset and the models. Finally, conclusions are drawn in Section VI.


\section{Dataset and Scenario}

The Smart Campus dataset \citep{realdata} is a research dataset aiming to encourage cooperation between academia and industry, promote the establishment of networks, facilitate the development of research-based campuses, and enable the testing of innovative Smart Campus services. 
This dataset finds applications in various areas, 
including  
time-series forecasting, anomaly detection, spatial correlation analysis, and occupancy estimation for specific areas. Furthermore, it is possible to apply various data analysis techniques to the LoRa parameters, such as battery consumption analysis, power analysis, and failure transmission analysis~\citep{eldeeb2023lorawan}. 

\begin{figure}[t!]
\centerline{\includegraphics[width=\columnwidth]{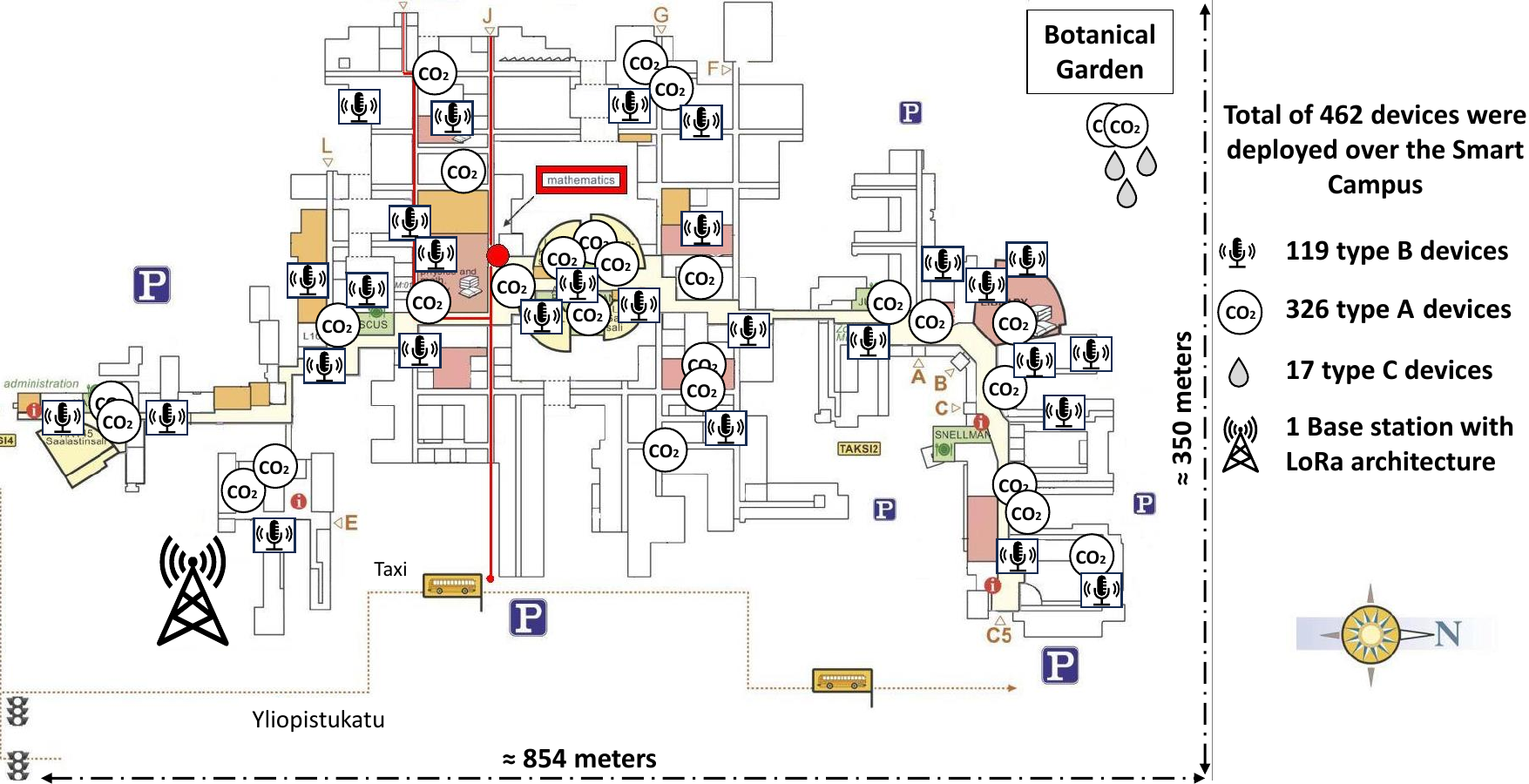}}
\caption{Map of the University of Oulu Smart Campus.}
\label{realUofO}
\end{figure}

The dataset comprises environmental information collected from 462 low-power sensors deployed across the University of Oulu campus and botanical garden. 
These sensors use Long-Range (LoRa) technology to transmit their readings to a base station (BS). 
The LoRa sensors are located across the $1.35 \times 10^5 \, \text{m}^2$ Linnanmaa campus and botanical garden (indoor and outdoor). 
The locations and types of sensors are represented in the online installation map, as depicted in Figure~\ref{realUofO}. 



\begin{figure}[t!]
\centerline{\includegraphics[width=0.6\columnwidth]{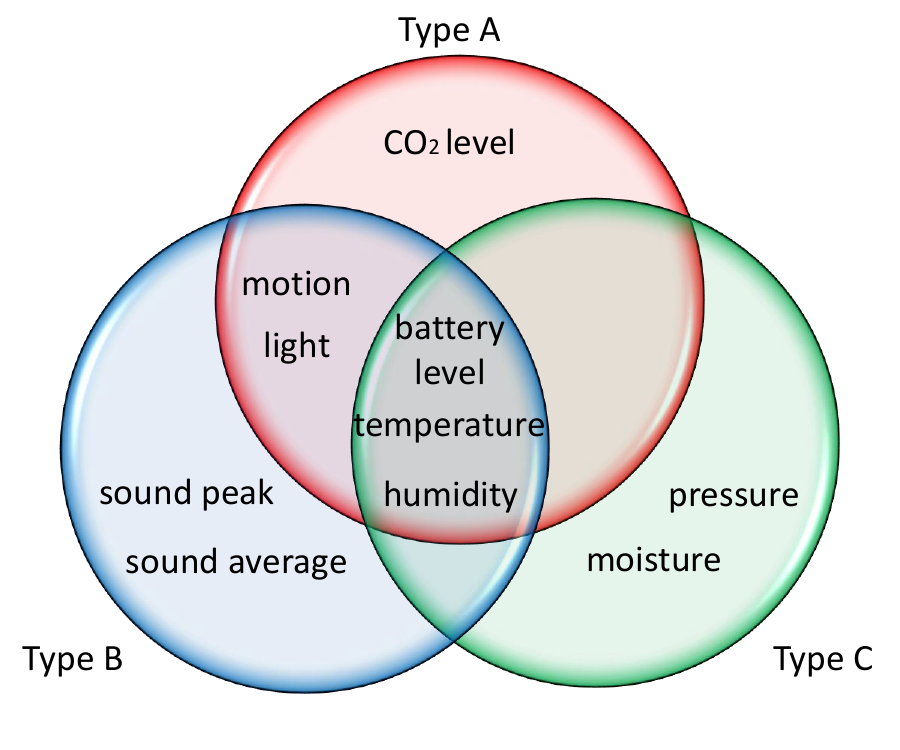}}
\caption{Types of sensors and their readings.}
\label{venn}
\end{figure}

The sensors in the Smart Campus dataset are categorized into three types: 
A, B and C. 
As depicted in Figure~\ref{venn}, type A devices measure CO$_2$ levels, motion, and light; type B measure sound average, sound peak, motion, and light; and type C devices measure pressure and moisture. Additionally, all devices track temperature, humidity, and battery levels. The dataset comprises of 462 devices, including 326 type A, 119 type B, and 17 type C devices. Each sensor is required to store its readings every 15 minutes and send them to the BS. In this scenario, 
the arriving traffic to the BS follows two different patterns depending on the monitored measures: 
(i) quasi-periodic or (ii) approximately event-driven (ED).

\section{Statistical analysis and Traffic models} 

The MTC traffic is usually uplink-dominated and characterized by short transmissions combining real-time and non-real-time traffic from multiple sources. According to its applications, MTC has three elementary traffic patterns~\citep{lopez2021csi}: (i) periodic update (PU), under which {devices transmit status reports regularly, \textit{e.g.}, smart-meter reading (gas, electricity, water)}; (ii) ED, which describes non-periodic traffic triggered by random events at an unknown time, \textit{e.g.}, alarms; and (iii) payload exchange (PE), which consists of {bursty traffic that usually comes after PU or ED traffic.} 

MTC traffic often appears as any of the aforementioned types of traffic or as a combinations of these types~\citep{IoT_FWuS}. For instance, an MTD may enter the power saving mode and trigger a PU pattern at regular intervals, while an alarm or critical event may activate the MTD and originate ED followed by PE traffic. Hence, using these three elementary classes enables building traffic models with different degrees of computational complexity and accuracy~\citep{eslam}.

Traffic modeling involves designing stochastic processes that mimic the behavior of measured data traffic. 
MTC traffic, as proposed in 3GPP, falls under the aggregated category. 
Nevertheless, even though MTC traffic exhibits aggregation characteristics, including some features of source traffic is beneficial to enhance the robustness and scalability of the modeling. 


In general, when analyzing the operation of communication networks servicing aggregated traffic, it is necessary to have appropriate mathematical models. In this regard, we aim to analyze different models to fit the traffic provided in the dataset. For this, we present four different approaches to model the different traffic patterns: (i) a PPP approach while considering event-driven traffic patterns with geometrically distributed burst duration, (ii) a quasi-periodic model, (iii) data augmentation from the original data, and (iv) curve fitting. Notice that the third approach needs the original data to recreate the data. Specifically, data augmentation involves generating additional training data by applying transformations like rotations, translations, scaling, and flipping to existing data. The aim of data augmentation is to expand the training dataset in terms of both size and diversity, ultimately enhancing the performance and resilience of ML models. This technique is particularly beneficial when dealing with small datasets~\citep{garcia2018survey}.  

Before delving into these modeling approaches, we review the 3GPP model in detail. The 3GPP model treats MTC as aggregated traffic. 
The 3GPP defines two distinct access demand profiles for MTC applications, referred to as Traffic models 1 and 2~\citep{3gpp2011study}. Traffic model 1 is characterized by a uniform probability, while traffic model 2 exhibits a burst-like probability modeled using the beta distribution with probability density function (PDF) given by
\begin{equation} \label{beta}
    f(t) = \frac{t^{\alpha-1}(T-t)^{\beta-1}}{T^{\alpha+\beta-1} \mathbf{B}(\alpha,\beta)}
\end{equation} within the time frame $0 \leq t \leq T$, where $\mathbf{B}$ designates the Beta function with parameters $\alpha,\beta > 0$. 

Now, let us discuss approaches (i) and (ii) to accurately model MTC traffic sources on a per-device basis. We will begin by explaining the system model, followed by an examination of event generation and its impact on nearby devices.

\subsection{System model}

\begin{figure}[t!]
	\centering
	\includegraphics[width=0.9\columnwidth]{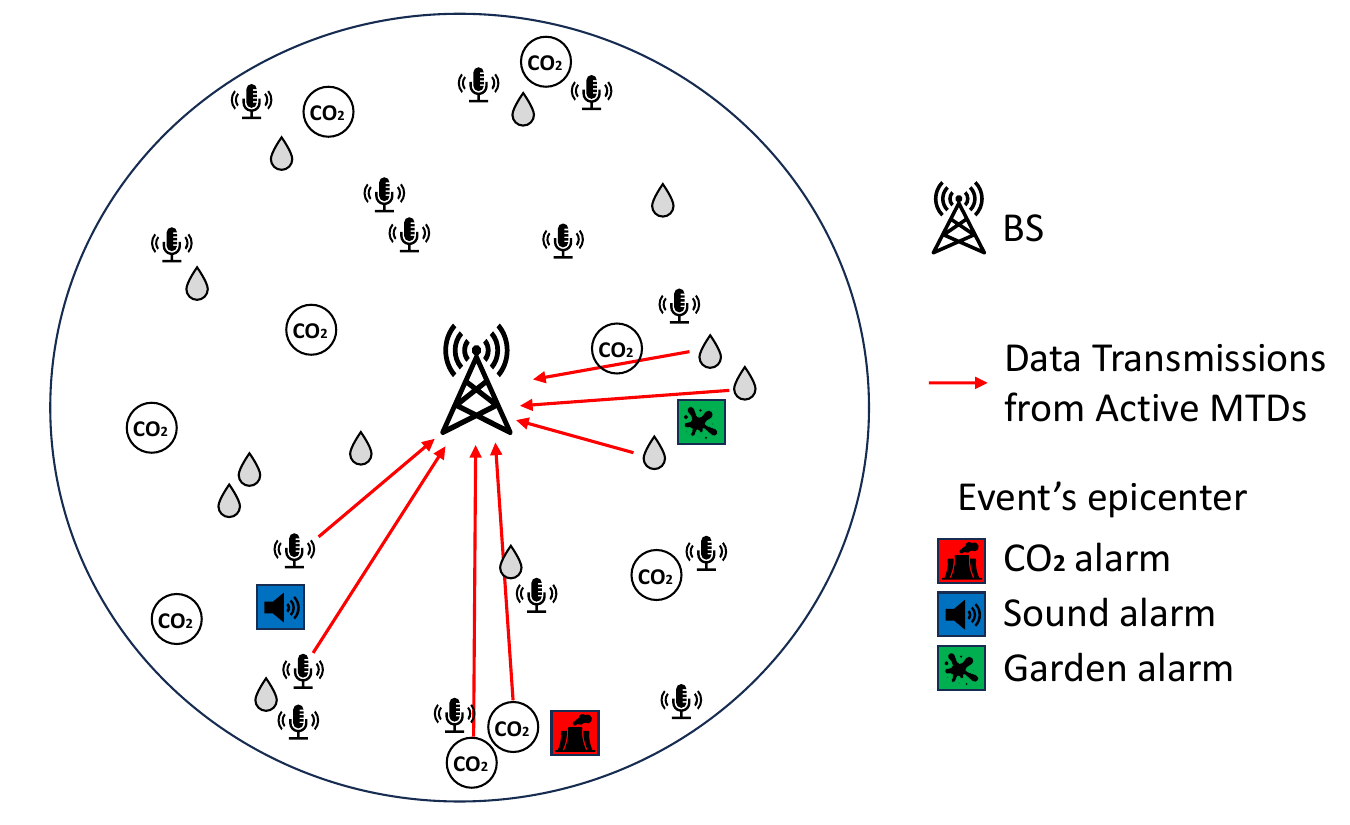}
	\caption{Illustration of an MTC network where a BS controls and collects information from the MTDs. 
 }
\label{figure1}
\end{figure}

We consider that the Smart Campus network can be abstractly represented as depicted in Figure~\ref{figure1}. The MTDs send packets to the BS, which controls all the information exchange within its cell. 
Each MTD can be idle (\textit{I}), waiting for a triggering event, or active (\textit{A}), exchanging information with the BS. The transition from state \textit{I} to \textit{A} occurs when information exchange between the MTD and BS is triggered due to the detection of an event. When the MTD goes to state \textit{A}, it stays there for the event duration. Assume that time is slotted in transmission time intervals (TTI). In 
state \textit{A}, the {MTD}s generate traffic with a rate $R$, while they do not generate traffic in state \textit{I}.

To model the position of MTDs and event epicenters, we use PPPs, as typical nodes and events can be reasonably assumed to be stochastically deployed in the Euclidean plane~\citep{IoT_FWuS}. Specifically, the MTDs are deployed according to a 2D homogeneous PPP $\Phi_{M}$ with density $\lambda_{M}$, while the event epicenters are represented by a 2D homogeneous PPP $\Phi_{E}$ with density $\lambda_{E}$. The processes $\Phi_{M}$ and $\Phi_{E}$ are assumed to be independent, while the BS is placed at the origin, as shown in Figure~\ref{figure1}.
\subsection{Influence of an Event Epicenter}\label{sec22}
To capture the effect of a given event on a sensing MTD, we define a function $p(d_{i,j})$ as the probability that an event in the $i^{th}$ epicenter ($i\in\Phi_E$) triggers the activation of an MTD \textit{j} at a location $j\in\Phi_M$, where $d_{i,j}$ is the distance between them in the Euclidean plane $\Re^{2}$~\citep{thomsen2017traffic}. Moreover, $p(d_{i,j}): [0, \infty) \rightarrow [0,1]$ is non-increasing to mimic a decaying influence of events as the distance $d_{i,j}$ increases~\citep{IoT_FWuS}. Figure~\ref{figure1} depicts the influence of an event epicenter on the surrounding MTDs.

\subsection{Generation of Events}\label{sec23}

Herein, we discuss two approaches to represent the event generation process in detail. Understanding the progression of events is crucial for MTC traffic analysis. 

\subsubsection{Poisson model}\label{sec231}

The time between the occurrence of events follows a Poisson distribution with density $\lambda_{T}$. Then, each MTD goes to state \textit{A} with probability~\citep{thomsen2017traffic}
\begin{equation}
	P_{A} = 1 - \exp\left({-2\pi\lambda_{T}\int^{\infty}_{0}{p(d)\partial d}}\right). 
    \label{PPPtraffic}	
\end{equation}

\subsubsection{Quasi-periodic traffic}\label{sec232}

Unlike periodic traffic, which is easy to predict once the periodicity parameters are found, quasi-periodic patterns are more challenging as they display irregular periodicity. This pattern is typical in industrial IoT scenarios~\citep{mitev2022smart}. Here, we consider MTC traffic characterized by homogeneous asynchronous periodicity. That is,  the transfer intervals $T_j$ for the $j^{th}$ MTD, defined as the number of time slots between consecutive transmissions, are independent and quasi-identically distributed~\citep{mitev2022smart}. The BS receives MTC signals at independent start times ($\kappa_j$). The activation probability and transmission duration for each MTD are denoted by $P_{A_j} \in [0,1]$ and $\delta_j$ respectively. The latter denotes the number of time slots required for each transmission. Finally, packets from a single MTD are transmitted with start time $t_{j}$ such that
\begin{equation}
    t_j = (\kappa_j + (m-1)T_j)B_j, \text{ } \hspace{2mm} m = 1, 2, \dots ,
    \label{quasiperiodic}
\end{equation}
where $B_j$ is a Bernoulli random variable with parameter $P_{A_j}$, and $m$ denotes 
the transmission opportunity. 
\subsection{Payload exchange}\label{sec24}
The PE patterns, whose durations are quantified by $\delta_j$, are modeled through the 
geometric distribution, where the parameter $q$ tunes the burstiness of the traffic generated by an event. Specifically, once in state \textit{A}, the MTD remains there for a number $k$ of TTIs with probability 
\begin{equation}
	{G_{k}} (q)= (1-q)q^{k}, \ k = 0,1,\dots
\label{geometric}	
\end{equation}
We assume that the value of $T_j$ is large enough so that the probability that two transmission opportunities overlap due to a relatively large $\delta_j$ ($k$ TTIs) is close to zero. Note that the parameter $q$ allows tuning the temporal correlation of the individual rate processes of the MTDs and, as a result, that of the total rate process, hence mimicking various MTC applications and event reporting strategies. For instance, in the case of a small $q$, the traffic behaves similarly to the one from a Bernoulli process (memoryless)~\citep{IoT_FWuS}. As $q$ increases, so does the memory since the total rate at a given time $k$ is correlated with past values. Then, after entering state \textit{A}, one remains there for longer~\citep{IoT_FWuS}. Notice that the traffic exchanged between the BS and the MTDs, when following a Poisson model, may be represented using an ergodic Markov chain~\citep{IoT_FWuS} with two states, \textit{I} and \textit{A}. 
\begin{figure}[t!]
	\centering
	\centerline{\includegraphics[width=0.4\columnwidth]{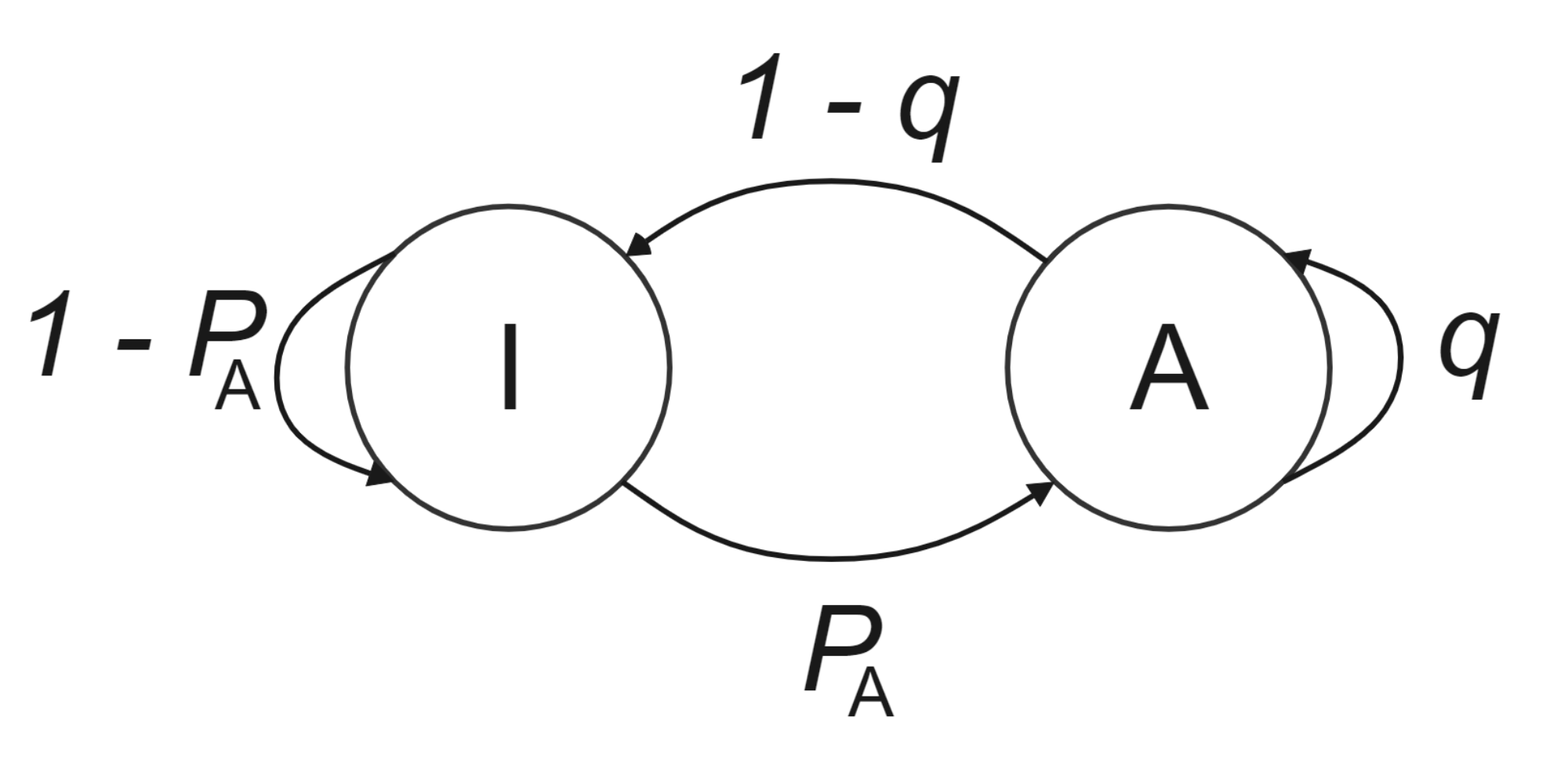}}
	\caption{Traffic exchanged between MTDs and the BS is modeled as a two-state complete Markov chain.}
\label{traffic}
\end{figure}

\begin{figure}[t!]
	\centering
        \centerline{\includegraphics[width=0.75\columnwidth]{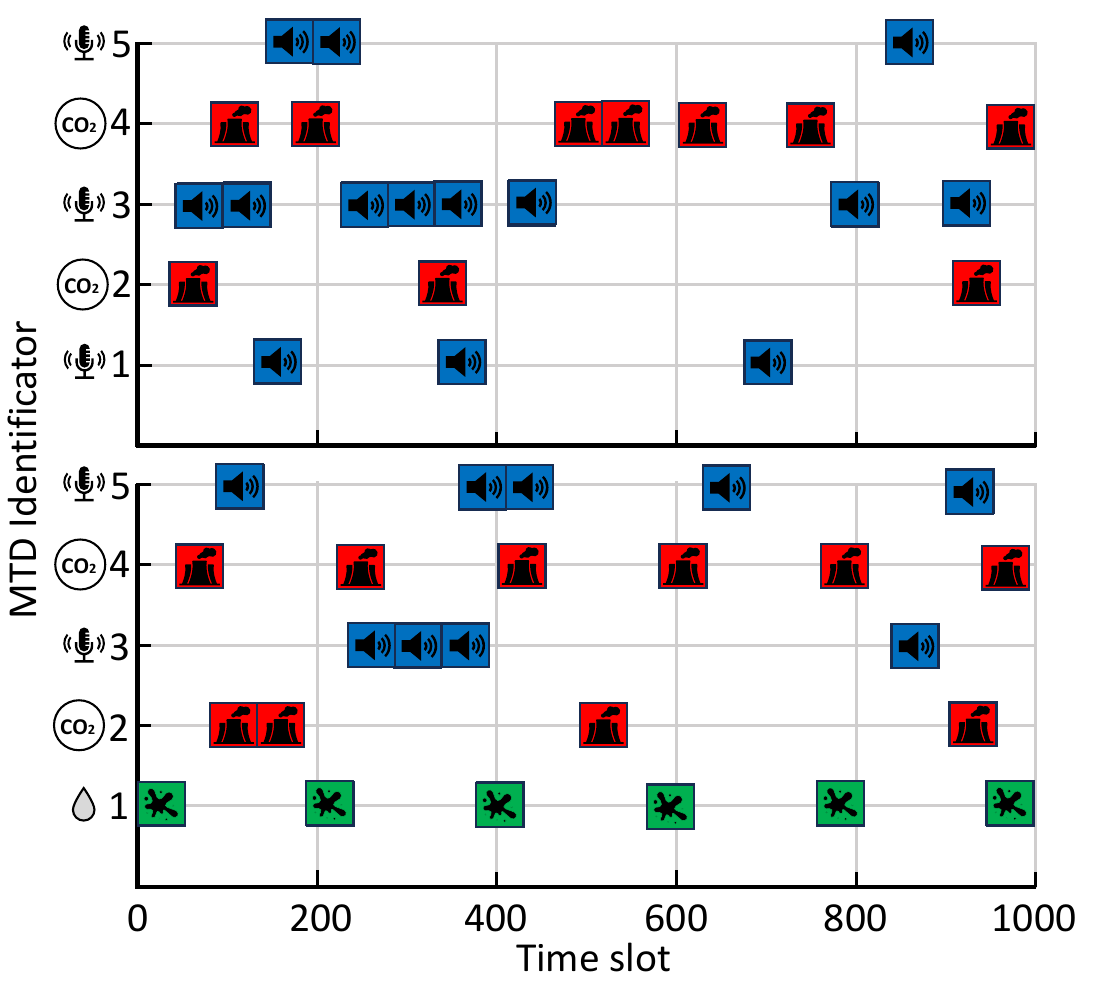}}
	\caption{Extract of a) Poisson (top) and b) quasi-periodic (bottom) traffic. The alarms represent the time slots (x-axis) where a given MTD (y-axis) is active.}
\label{quasi_traffic}
\end{figure}

Figure~\ref{traffic} describes the traffic exchanged between each MTD and the BS as a two-state semi-Markov chain, while extracts from the approach models (i) and (ii) illustrating the traffic pattern generated from the MTDs are depicted in Figure~\ref{quasi_traffic}. 
More specifically, the diagram illustrates traffic patterns from 5 MTDs. The top section displays Poisson patterns, while the bottom part showcases quasi-periodic patterns. 
The alarms represent the time slots where a given MTD is active. 
Some MTDs periodically have traffic patterns, while others are rarely active due to unequal activation probabilities. Furthermore, distinct burstiness patterns are visible.

\section{Performance metrics}

In this section, we discuss RMSE , the K-S test, the A-D test, and the C-S test, {which are} utilized later to assess the performance of the models and compare them with the real data. These statistics evaluate the effectiveness of the fit in accounting for the data's variability, indicating the degree of correlation between the model and the actual values.

\subsection{RMSE}

We calculate the RMSE between the arriving traffic using the models ($g_{i}$) and the actual real data ($\hat{g}$) values as
\begin{equation}
	\text{RMSE} = \sqrt{{\frac{1}{M}\sum_{i=1}^{M}\left(g_{i}-\hat{g}\right)^{2}}},
\label{eq6}	
\end{equation}
{where $M$} is the sample size.

\subsection{Kolmogorov-Smirnov test}

The K-S test is a statistical test used to determine if two samples are drawn from the same distribution by comparing their cumulative distribution functions (CDFs)~\citep{massey1951kolmogorov}. 
It provides the maximum vertical deviation (D statistic) between the empirical distribution function of the sample data and the theoretical cumulative distribution function. 
The test can be applied to any sample set but is most commonly used for continuous distributions. The null hypothesis is that the two samples come from the same distribution, and the alternative hypothesis is that they come from different distributions~\citep{berger2014kolmogorov}. 
The test is easy to perform and interpret but has limitations, such as insensitivity to differences beyond the CDF and lack of information on the nature of differences. The K-S test is particularly useful for assessing the normality of data.

\subsection{Anderson-Darling test}

The A-D test of goodness-of-fit is a statistical tool used to evaluate how well a given dataset conforms to a particular theoretical distribution \citep{shao2023research}. It extends the capabilities of the K-S test, especially in assessing the fit of data in the tails of the distribution. The test works by comparing the observed data to expected values derived from the assumed distribution, calculating a statistic that reflects the degree of fit. This statistic is then compared to critical values, typically obtained from statistical tables or software, to determine whether the data significantly deviates from the chosen theoretical distribution. A higher A-D statistic suggests a poorer fit, while a lower value indicates a better fit. The A-D test is a valuable tool for assessing data conformity to specific distributions \citep{samuh2023distribution}.

\subsection{Chi-squared test}

The C-S test is a statistical method commonly used to assess the association or independence between categorical variables. It works by comparing observed frequencies in a contingency table to the expected frequencies if there were no association between the variables. The test involves checking the discrepancies between observed and expected frequencies. 
The C-S test is widely used to determine if there is a significant relationship between categorical variables, making it a valuable tool for hypothesis testing and data analysis \citep{chi}. However, this test gives only evidence of an association or no association, and it does not produce effect estimates and confidence intervals. In other words, it does not tell us the effect (risk ratio or odds ratio). In addition, the application of the C-S test highly depends on the sample size and the magnitude of the expected frequencies.

\section{Analysis Results}

In this section, we compare the data collected during 11 months from \citep{realdata} and those generated by the models considered in this paper. First, we process and classify the raw data and divided the data according to their correspondence to quasi-periodic and non-periodic traffic using the timestamp in the dataset. Herein, measurements packets including sound or CO$_2$ levels present non-periodic patterns while others like temperature or moisture levels show quasi-periodic ones. Then, we aim to use fitting techniques to adjust the most suitable well-known models for the data.  

\subsection{Fitting} 

We first fit the data using Matlab \citep{Matlab} and EasyFit \citep{mathwave2015easyfit} software, to obtain a ranking for well-known models. Table~\ref{fit} reflects the results for the goodness of fit using three different statistical tests to analyze the data: (i) K-S test~\citep{KS_table,samuh2023distribution}; (ii) A-D test~\citep{shao2023research,samuh2023distribution}
; and (iii) the C-S test~\citep{chi}
. All tests were made for significance level ($\alpha = 0.01$) to make decisions about the null hypothesis. This significance level represents the probability of mistakenly rejecting the null hypothesis when, in fact, it is accurate. The statistics provided by K-S and A-D in the table allow for the ranking of each distribution based on their goodness-of-fit. Additionally, the tests that have passed the C-S test are highlighted in red for easy identification.  

\begin{table}[t!]
\caption{Goodness of Fit Tests. The tests that have passed the C-S test are indeed highlighted in red.}
\begin{center}
\begin{tabular}{clclc} 
\toprule
\textbf{Rank} &{\textbf{Model}}&\textbf{K-S Statistic}&{\textbf{Model}} &\textbf{A-D Statistic}\\
\midrule 
1& \cellcolor{red!25} {Gen. Pareto}&0.083 &\cellcolor{red!25} {Gen. Pareto}&0.117\\
2&\cellcolor{red!25} {Beta}&0.091 &\cellcolor{red!25} {Beta}&0.122\\
3&Johnson SB&0.120 &\cellcolor{red!25} {GEV$^{\mathrm{a}}$} &0.157\\
4&\cellcolor{red!25} {Pearson 6}&0.121 &Johnson SB &0.162\\
5& Weibull&0.137 &Weibull&0.198\\
\bottomrule
\multicolumn{3}{l}{$^{\mathrm{a}}$Generalized Extreme Value (GEV).}
\end{tabular}
\label{fit}
\end{center}
\end{table}

From Table~\ref{fit}, we can see that the Generalized Pareto~\citep{haj2022generating} with PDF given by 
\begin{equation}
    \begin{array}{cl}
        f(x;\theta,\lambda) &= \begin{cases}
                        \frac{1}{\lambda}\left(1+\frac{\theta}{\lambda}x\right)^{(1+\frac{1}{\theta})},   &\theta \neq 0,\\
                        \frac{1}{\lambda}\exp{(-\frac{x}{\lambda})},   &\theta = 0,\\
                        \end{cases}\\
        \end{array}
\end{equation}
and the Beta distributions (as proposed by 3GPP, \eqref{beta}) 
provide the best fit to the data and  successfully passed the three goodness-of-fit tests. Then, we use these distributions as benchmarks for the model presented before. 

\subsection{Comparison}


In this section, we focus on analyzing and evaluating different traffic models to identify the ones that best characterize the traffic patterns observed in MTC scenarios. The aim is to determine which traffic models 
accurately characterize the MTC traffic, thus favoring research based on 
simulations and traffic predictions of MTC scenarios. By comparing and contrasting these traffic models, the study seeks to provide valuable insights into the suitability and performance of each model in capturing the unique characteristics of MTC traffic patterns.

\subsubsection{Event-driven data}

\begin{figure}[t!]
            \centering
            \centerline{\includegraphics[width=\columnwidth]{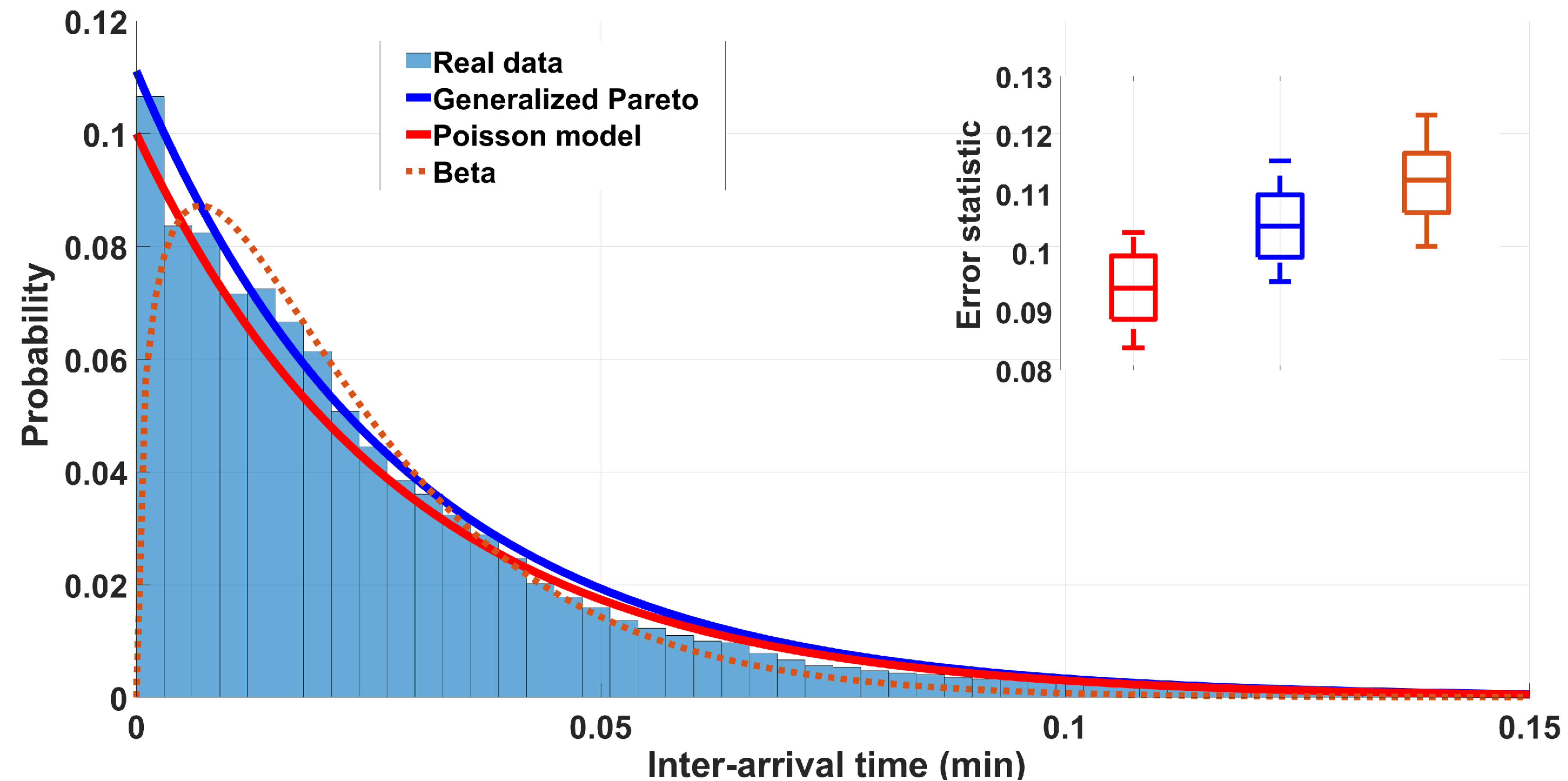}}
            \vspace{-2mm}
            \caption{{Histogram of event-driven real data and the Poisson model.}}
            \label{real}
        \end{figure}

Figure~\ref{real} 
shows the inter-arrival times distribution related to event-driven traffic considering real data and synthetically generated data. 
In event-driven data, the information is collected or updated only 
upon the occurrence of predefined events rather 
than being continuously sampled at fixed intervals. 
Specifically, the histograms are associated with the sound and CO$_2$ measures in the data packets collected from the sensor network at the Smart Campus at the University of Oulu. 
Conversely, the curves represent the fitting 
of 
the inter-arrival time data to the models using the traffic statistics from the real data. We can conclude from the figure that the models characterize pretty accurately the real data with an estimation error smaller than 11\% (goodness-of-fit tests) and an RMSE of 0.947.
Additionally, the error in the test shows the goodness of fit regarding each model. Specifically, 
we can see how the proposed PPP fits better the data, followed by the generalized Pareto and the Beta distribution.

\begin{figure}[t!]
            \centering
            \centerline{\includegraphics[width=\columnwidth]{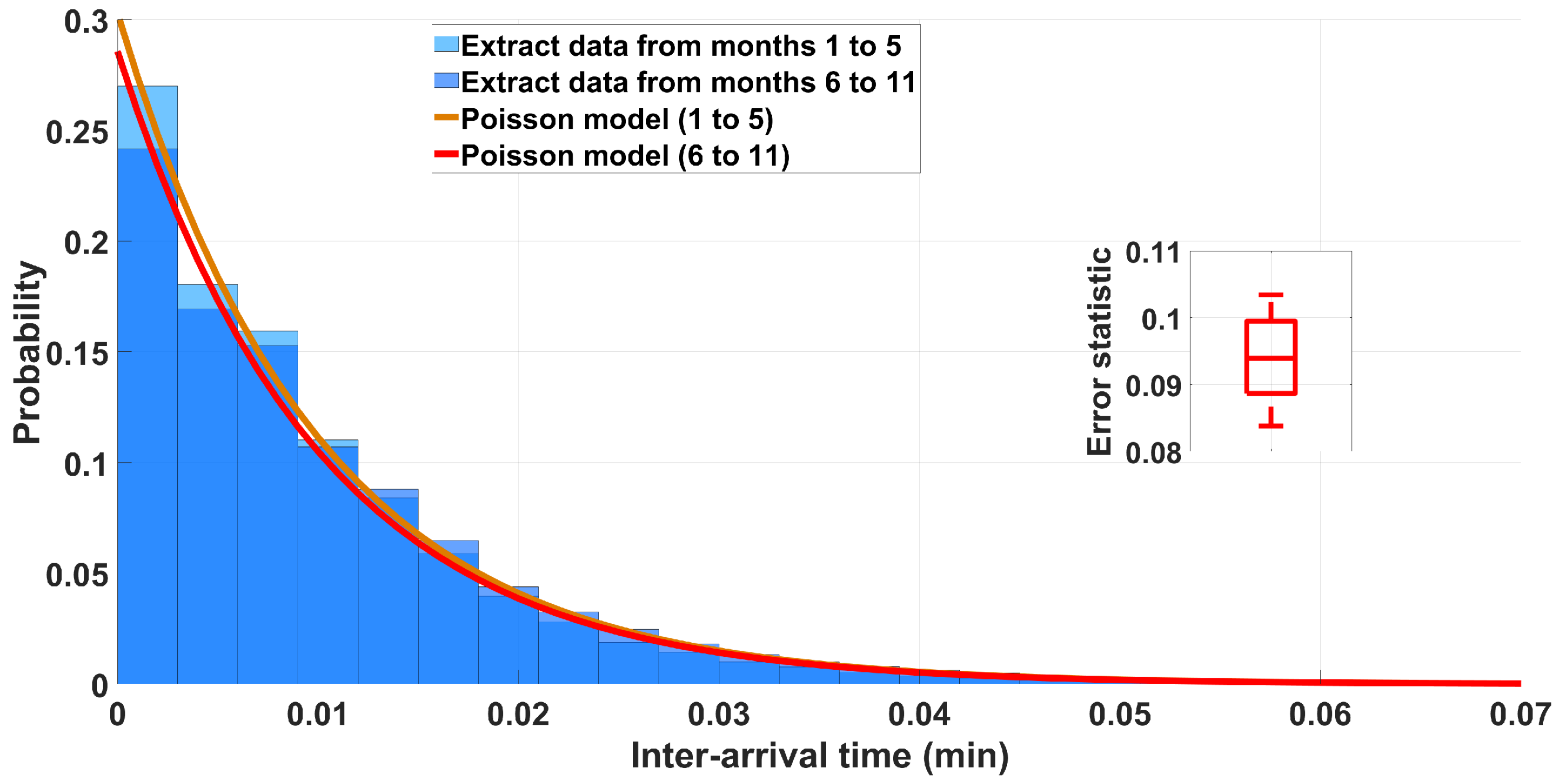}}
            \vspace{-2mm}
            \caption{{Event-driven real data during different measurement intervals.}}
            \label{6m}
        \end{figure}

Note that the histogram for the event-driven real data is split into two parts comprising half a year each. From Figure~\ref{6m}, we can see that the trends in both periods follow similar patterns, ratifying the memoryless property of this kind of traffic.
Nevertheless, in general, findings show a high correlation between the real data and the model-generated one. 

\subsubsection{Quasi-periodic data}

\begin{figure}[t!]
            \centering
            \centerline{\includegraphics[width=\columnwidth]{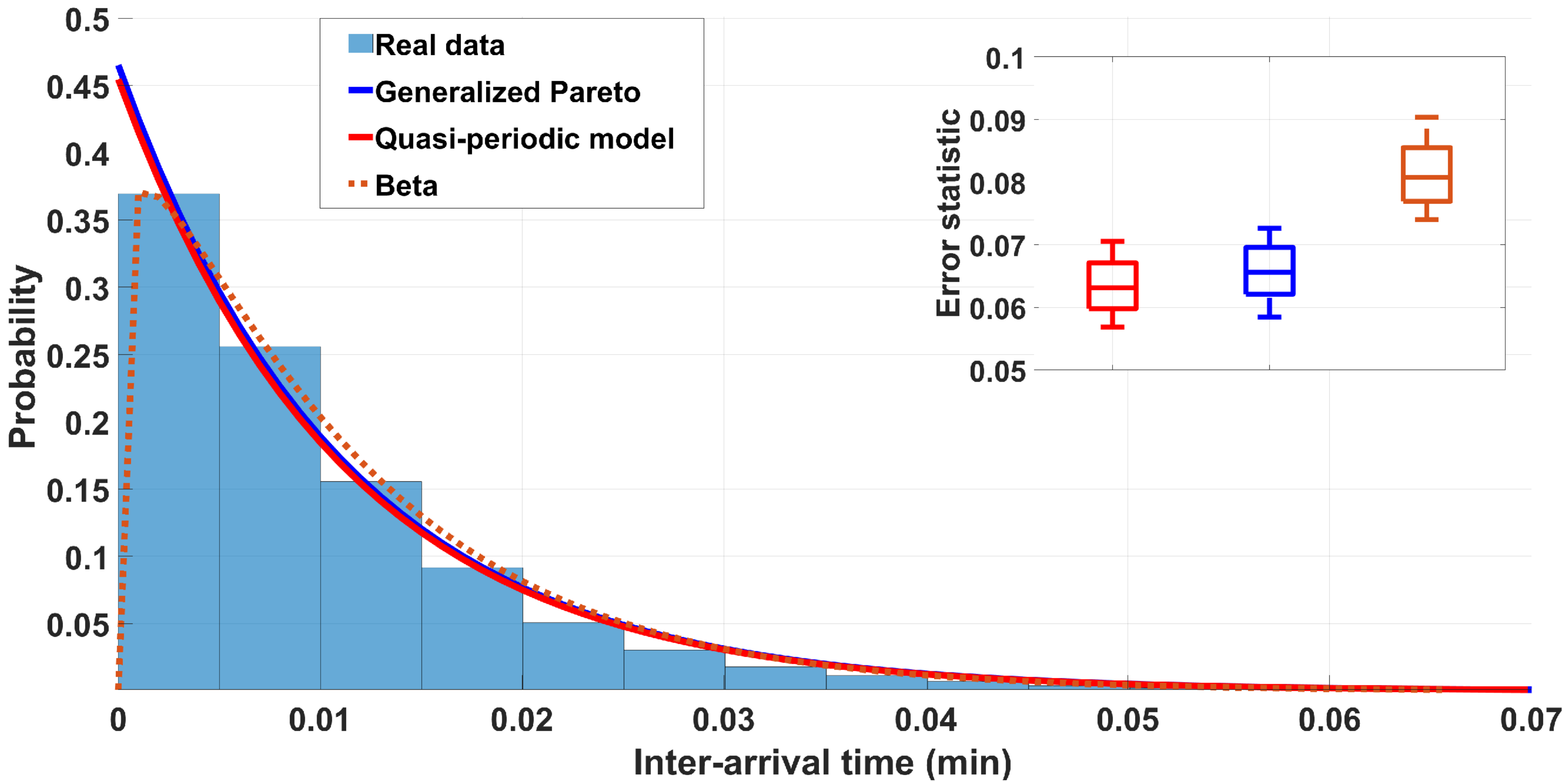}}
            \vspace{-2mm}
            \caption{{Histogram of quasi-periodic real data and model.}}
            \label{realq}
        \end{figure}

Figure~\ref{realq} shows the distribution of the inter-arrival times for both, real and synthetically-generated data, considering now the quasi-periodic traffic. The histograms are associated with the temperature measures in the data packets collected from the sensor network at the Smart Campus at the University of Oulu, while the curves represent the fitting of the inter-arrival time data for the models, as described before for event-driven data. We can see from the figures that the models characterize the real data with an estimation error smaller than 7\% and a RMSE of 0.982.    
The quasi-periodic model proposed in this paper fits better the real data 
compared with 
the generalized Pareto and beta models attaining 7.5\% and 9\% estimation errors, respectively. 

\subsubsection{Extreme Event Probabilities}

\begin{figure}[t!]
    \centering
    {\includegraphics[width=0.49\linewidth]{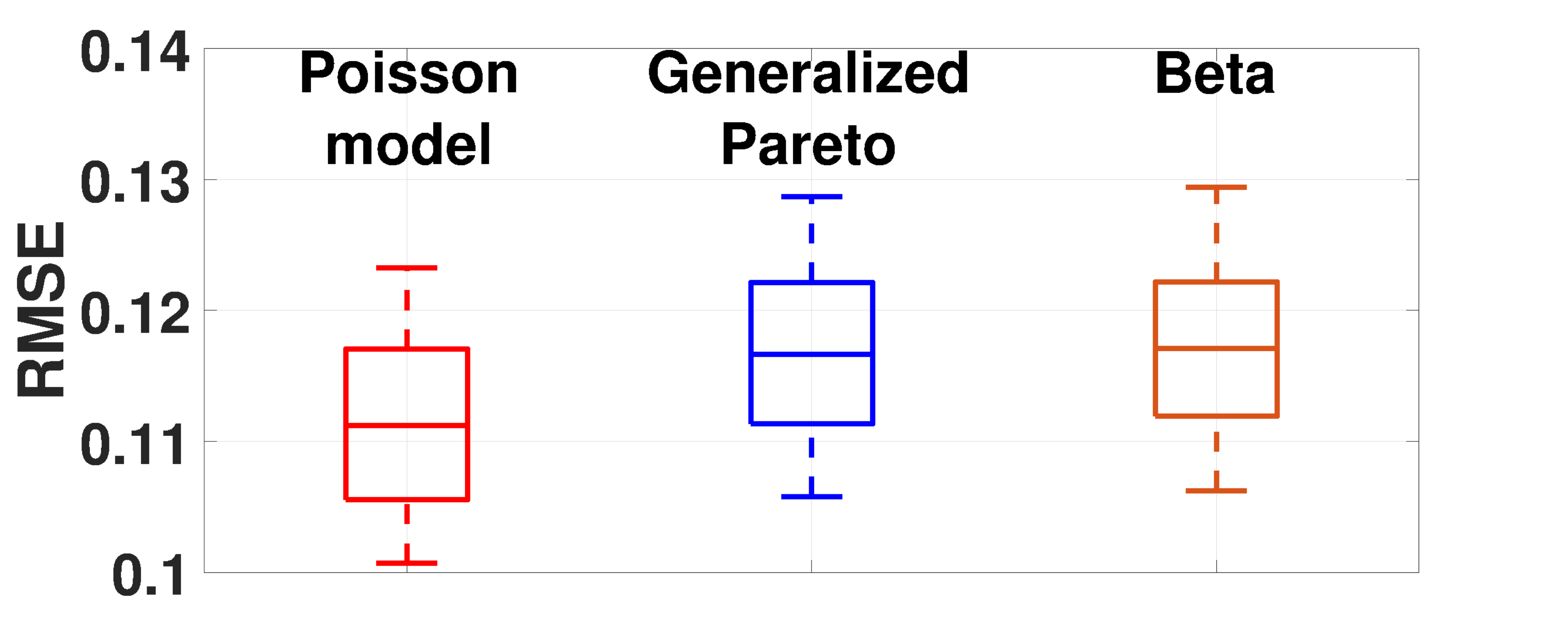}}
    {\includegraphics[width=0.49\linewidth]{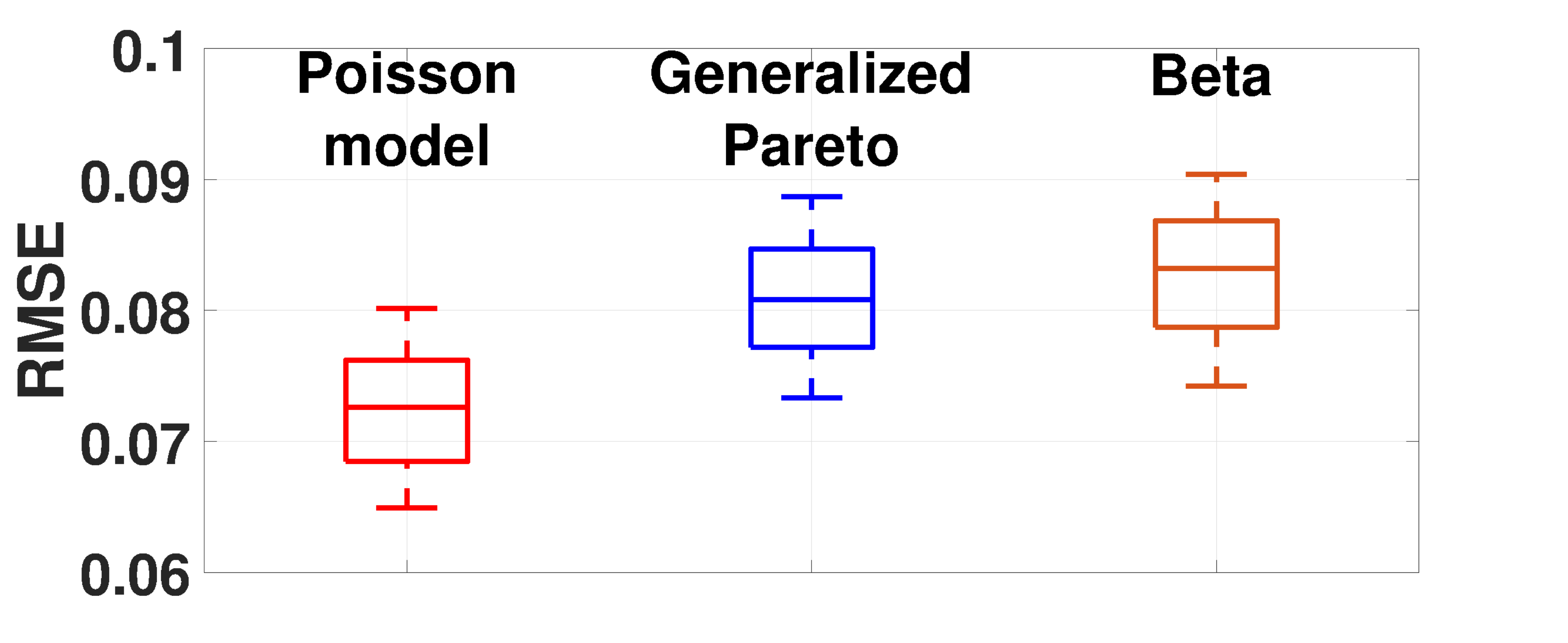}}
    \vspace{-2mm}
    \caption{{Prediction error at the tails of the models' function for a) Poisson (left) and b) quasi-periodic (right) traffic.}}
    \label{tail}
\end{figure}

Additionally, we illustrate the models' error in the tails, \textit{i.e.}, for modeling minimum and maximum inter-arrival time in Figure~\ref{tail}. Note that comparing the cumulative distribution functions (CDFs) tails is helpful to assess the behavior of extreme traffic patterns and the likelihood of encountering unusual data transmission events. This analysis is essential for understanding the impact of rare but critical occurrences in MTC, which can have implications for network performance, resource allocation, and overall system reliability. By comparing the CDFs tails in MTC traffic modeling, one can gain insights into how well the model represents the extreme traffic scenarios, ensuring more accurate predictions and better decision-making in managing MTC networks. 

In Figure~\ref{tail}, the error probabilities at the tails (below 5\% and above 95\% of the cases) of the previously analyzed models are presented. Notably, the proposed PPP and quasi-periodic models maintain the lowest level of error at the tails, with errors measuring below 0.12 and 0.08, respectively. In contrast, the generalized Pareto and beta models exhibit a relatively similar level of error, approximately 0.13 for ED traffic patterns and 0.09 for PU traffic patterns.


\section{Conclusions}

This paper addressed the gap in understanding MTC traffic patterns by characterizing the Smart Campus dataset from the University of Oulu. A statistical analysis, including the Kolmogorov-Smirnov test, was carried out to study the two most significant types of traffic in MTC: periodic updating and event-driven. The paper compared and evaluated various traffic models suitable for simulating real-world scenarios corresponding to each traffic type. The accuracy of the models was assessed through a comprehensive analysis of their goodness-of-fit, using tests such as K-S, A-D, C-S, and RMSE. 
The results showed that the Poisson point process model provides the best fit for event-driven patterns, with errors below 11\%, while the quasi-periodic model exhibits the best fit for periodic updating, with errors below 7\%. This analysis sheds light on MTC traffic behavior and contributes to a better understanding of the data.
It was demonstrated that the proposed models can be used with a high level of confidence for MTC traffic modeling, simulation, and prediction. 

In future research, we intend to explore scenarios where the network operates in a saturated regime and with  transmission cycles leading to increased collision rates. These factors play a crucial role for a more complete comprehension of MTC traffic dynamics.

\section*{Acknowledgements}
This work is partially supported by the Finnish Foundation for Technology Promotion and the Research Council of Finland (former Academy of Finland) 6G Flagship Programme (Grant Number: 346208), as well as in Chile by the ANID FONDECYT Regular. We want to express our gratitude to Aleksi Pirttimaa (Assistant Researcher, Smart Campus Dataset curator) for their valuable assistance during this research. 

\bibliography{bib}

\end{document}